\documentclass[12pt,a4paper]{article}

\usepackage[margin=1in]{geometry}
\usepackage{setspace}
\usepackage{lmodern}
\usepackage{amsmath,amssymb,amsfonts,bm}
\usepackage{braket}

\usepackage{physics}

\usepackage{graphicx}
\usepackage{subcaption}
\usepackage{float}

\usepackage[numbers,sort&compress]{natbib}
\usepackage[hyphens]{url}
\usepackage[hidelinks]{hyperref}
\usepackage{microtype}

\usepackage{authblk}

\title{\textbf{A Quantum Circuit Model of Black Hole Evaporation with Tunable Semi-Causality Violation}}

\author[1,2]{Sourav Ballav\thanks{\texttt{souravballav@gmail}}}
\author[1,3]{Wen-Yu Wen\thanks{\texttt{wenw@cycu.edu.tw}}}
\author[1]{Chi-Hsien Tai\thanks{\texttt{xkp92214@gmail.com}}}

\affil[1]{Department of Physics and Center for Theoretical Physics\\
Chung Yuan Christian University, Taoyuan 32023, Taiwan}

\affil[2]{Physics Division, National Center for Theoretical Sciences\\
Taipei 10617, Taiwan}

\affil[3]{Leung Center for Cosmology and Particle Astrophysics\\
National Taiwan University, Taipei 10617, Taiwan}

\date{}

\begin{document}

\maketitle

\begin{abstract}
We present a four-qubit quantum circuit model of black hole evaporation with a controlled violation of semi-causality, understood as the condition that information may fall into the black hole but cannot propagate back across the horizon. Building on Broda's semi-causal evaporation circuit, we introduce a controlled-unitary gate $\mathrm{CU}(\sigma)$ that allows tunable information leakage from the interior to the exterior while preserving global unitarity.

We compute the single-qubit reduced entropies, together with the mutual
information and entanglement negativity for the BH-GR and IN-OUT
bipartitions, at each discrete time step. For $\sigma=0$, the model
reproduces Broda's Page-like entropy evolution with complete late-time
purification. For any $\sigma>0$, however, nonzero residual single-qubit
entropies and persistent late-time entanglement remain, indicating
incomplete purification of the outgoing radiation despite the global
unitary evolution.

In the small-$\sigma$ regime, the residual entropy exhibits a characteristic $-\sigma^{2}\ln\sigma^{2}$ scaling that bears qualitative similarity to logarithmic entropy corrections in generalized uncertainty principle inspired evaporation scenarios. For larger values of $\sigma$, the persistence of finite residual entropy invites comparison with remnant-like endpoint configurations in regular or extremal black hole models. Our results show how controlled departures from the semi-causal limit modify information recovery in a minimal analytically tractable model of black hole evaporation.

\end{abstract}

\section{Introduction}
\label{sec:intro}

Hawking's discovery of black hole radiation \cite{Hawking:1974sw,Hawking:1976ra}
revealed a fundamental tension between general relativity and quantum mechanics. If black hole evaporation is exactly thermal and proceeds to completion, an initially pure quantum state evolves into a mixed state, apparently violating
unitary time evolution. This tension underlies the black hole information paradox, which may be formulated in terms of the fine-grained entropy of Hawking radiation and the fate of quantum correlations across the event horizon
\cite{Preskill:1992tc}.

Significant recent progress has been made through semiclassical gravitational path integrals and the island formula \cite{Penington:2019npb,Almheiri:2019psf,Almheiri:2019hni,Geng:2024xpj,Geng:2025efs}, which provide replica-wormhole saddle points and recover a Page-like entropy curve for the radiation~\cite{Page:1993wv,Page:2013dx}. These developments, together with quantum extremal surfaces \cite{Akers:2021fut}, suggest that unitarity can be preserved provided semiclassical gravity is suitably corrected near or at the horizon. Parallel ideas from quantum error correction and holography \cite{Harlow:2016vwg,Almheiri:2014lwa} emphasize that interior–exterior entanglement can be encoded redundantly in the radiation degrees of freedom.

In addition to semiclassical approaches, quantum circuit models provide a complementary toy framework in which evaporation can be represented by a unitary time evolution on a finite set of qubits \cite{Avery:2013exa,Broda:2021gts,Tokusumi:2018typ, Lu:2026qbg,Yang:2020riy,Osuga:2016htn,Verlinde:2021kgt,Piroli:2020dlx,Broda:2019ofl}. These models track the distribution of entanglement directly and allow discrete computation of von Neumann entropy, mutual information, and negativity between interior and radiation sectors. Broda \cite{Broda:2021gts} introduced a minimal four-qubit evaporation circuit and implemented a constraint known as semi-causality: information may flow from exterior qubits into the black hole interior, but not from interior to exterior across the horizon\cite{Broda:2023gxs}. Because of its simplicity, the model permits an essentially analytic treatment of the evaporation process while reproducing a Page-like entropy curve. In the language of quantum information theory \cite{EggelingWerner,SchumacherWestmoreland,BeckmanGottesman}, semi-causality is a one-way signalling condition on bipartite quantum operations, and leads in Broda’s model to a Page-like entropy evolution with complete late-time purification of the radiation.

From a quantum gravity perspective, it is natural to ask whether the strict semi-causal constraint imposed in such circuit models remains exact. Related modifications of late-time black hole evaporation have been discussed in scenarios involving generalized uncertainty principle (GUP) corrections, regular or extremal black holes, and remnant formation. However, the consequences of controlled semi-causality violations for entanglement structure have not been systematically explored in circuit-based models.

In this work we develop a four-qubit quantum circuit model of black hole evaporation that incorporates a tunable violation of semi-causality. We introduce an interior-exterior controlled-unitary gate $\mathrm{CU}(\sigma)$, where the parameter $\sigma$ continuously interpolates between Broda's strictly semi-causal circuit $(\sigma=0)$ and a regime of maximal information leakage within the present circuit architecture $(\sigma=\pi/4)$. Unlike previous circuit models that recover complete late-time purification, the controlled-unitary deformation introduced here generates finite residual entropy and persistent entanglement while preserving global unitarity.

We compute von Neumann entropy, mutual information, and entanglement negativity at each time step and show that any $\sigma>0$ leads to nonzero residual entropy and persistent late-time entanglement across the horizon. For small $\sigma$, the resulting entropy plateau exhibits a characteristic $-\sigma^{2}\ln\sigma^{2}$ scaling that bears qualitative similarity to logarithmic entropy corrections in GUP-inspired evaporation scenarios. For larger values of $\sigma$, the persistence of finite residual entropy invites comparison with remnant-like endpoint configurations in regular or extremal black hole models. Our circuit thus provides a minimal analytically tractable framework for exploring how controlled departures from semi-causality modify information recovery during black hole evaporation.

\section{Quantum circuit model with controlled semi-causality violation}
\label{section2}

Causality, the principle that no information can propagate faster than light, is a foundational aspect of classical general relativity. In the context of black hole evaporation, enforcing causality typically implies that quantum information cannot escape from behind the event horizon. Several early qubit-based evaporation models~\cite{Giddings:2012bm,Mathur:2011wg,Osuga:2016htn,Avery:2013exa} incorporated unitary evolution but did not explicitly impose causal constraints across the horizon. 

Broda's four-qubit circuit model~\cite{Broda:2021gts} addressed this by building causal structure directly into the gate dynamics: information was permitted to flow from the near-horizon modes into the black hole interior but not in the reverse direction, a condition termed \emph{semi-causality}. In particular, only CNOT gates with \emph{external} control (GR, OUT) and \emph{internal} target (BH) were allowed across the horizon, while SWAP operations across the horizon and CNOTs in the opposite direction were forbidden. This scheme reproduced a Page-like entropy curve under globally unitary evolution while prohibiting any leakage of information outward from the interior.

In this work, we preserve Broda's architecture but introduce a controlled and continuously tunable violation of semi-causality through a single-parameter controlled-unitary gate, $\mathrm{CU}(\sigma)$. Unlike the original model, which enforces strict semi-causality, our construction allows a continuous interpolation between the semi-causal limit and a regime of maximal information leakage within the present circuit architecture while maintaining global unitarity. This simple modification provides a minimal framework for studying how controlled departures from semi-causal evolution affect the entanglement structure of black hole evaporation. The parameter $\sigma$ quantifies the degree of semi-causality violation and serves as the central control parameter throughout our analysis.

\begin{figure}[t]
\centering
\includegraphics[width=0.55\textwidth]{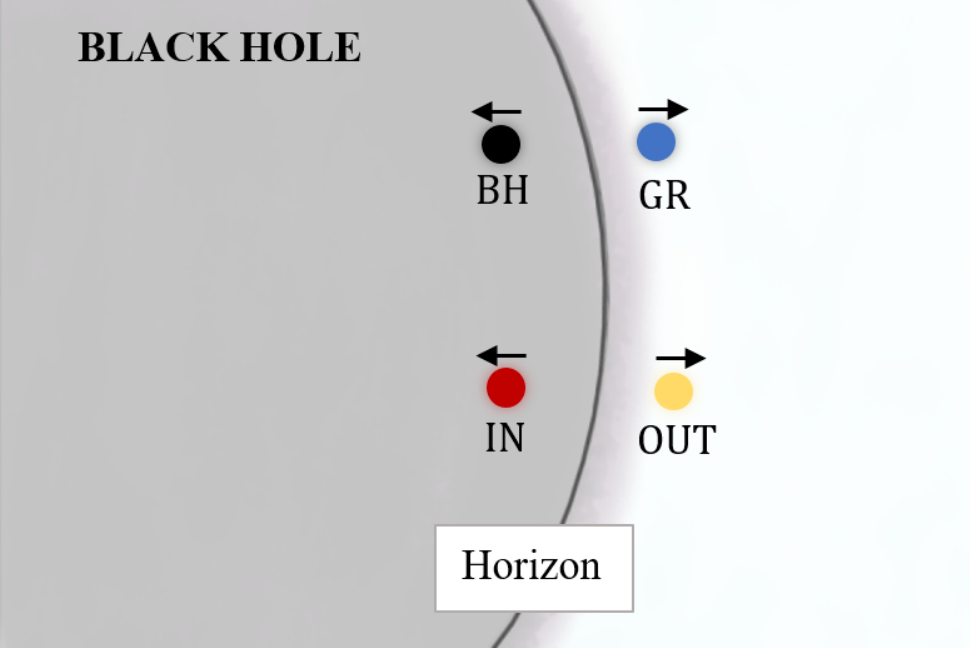}
\caption{Pictorial representation of the qubit model for black hole evaporation. The four qubits correspond to BH (interior), GR (near-horizon gravity), IN (infalling partner), and OUT (Hawking radiation).}
\label{figBH}
\end{figure}

We now describe our model of black hole evaporation, which is schematically illustrated in Fig.~\ref{figBH}.

\subsection{Circuit setup}
\label{subsec:setup}

The quantum circuit consists of four qubits representing the key subsystems involved in black hole evaporation:
\begin{itemize}
  \item \textbf{BH}: black hole interior (collapsing matter),
  \item \textbf{GR}: near-horizon gravitational degrees of freedom,
  \item \textbf{IN}: the infalling Hawking partner mode,
  \item \textbf{OUT}: the outgoing Hawking radiation.
\end{itemize}

The total Hilbert space is
\begin{equation}
\mathcal{H}_{\mathrm{tot}}
= \mathcal{H}_{\mathrm{BH}} \otimes \mathcal{H}_{\mathrm{GR}}
  \otimes \mathcal{H}_{\mathrm{IN}} \otimes \mathcal{H}_{\mathrm{OUT}},
\end{equation}
where each subspace is two-dimensional, so that 
\(\dim \mathcal{H}_{\mathrm{tot}} = 2^4 = 16\).

The system is initialized in the pure product state
\begin{equation}
\ket{\Psi(0)}
= (\alpha \ket{0} + \beta \ket{1})_{\mathrm{BH}}
  \otimes \ket{0}_{\mathrm{GR}}
  \otimes \ket{0}_{\mathrm{IN}}
  \otimes \ket{0}_{\mathrm{OUT}},
\qquad |\alpha|^2 + |\beta|^2 = 1.
\end{equation}
Here $\ket{0}$ represents the vacuum state for each qubit. The black hole qubit 
$\ket{\xi}_{\mathrm{BH}} = \alpha \ket{0} + \beta \ket{1}$
encodes the initial information content of the collapsing matter.

\begin{figure}[t]
\centering
\includegraphics[width=0.95\textwidth]{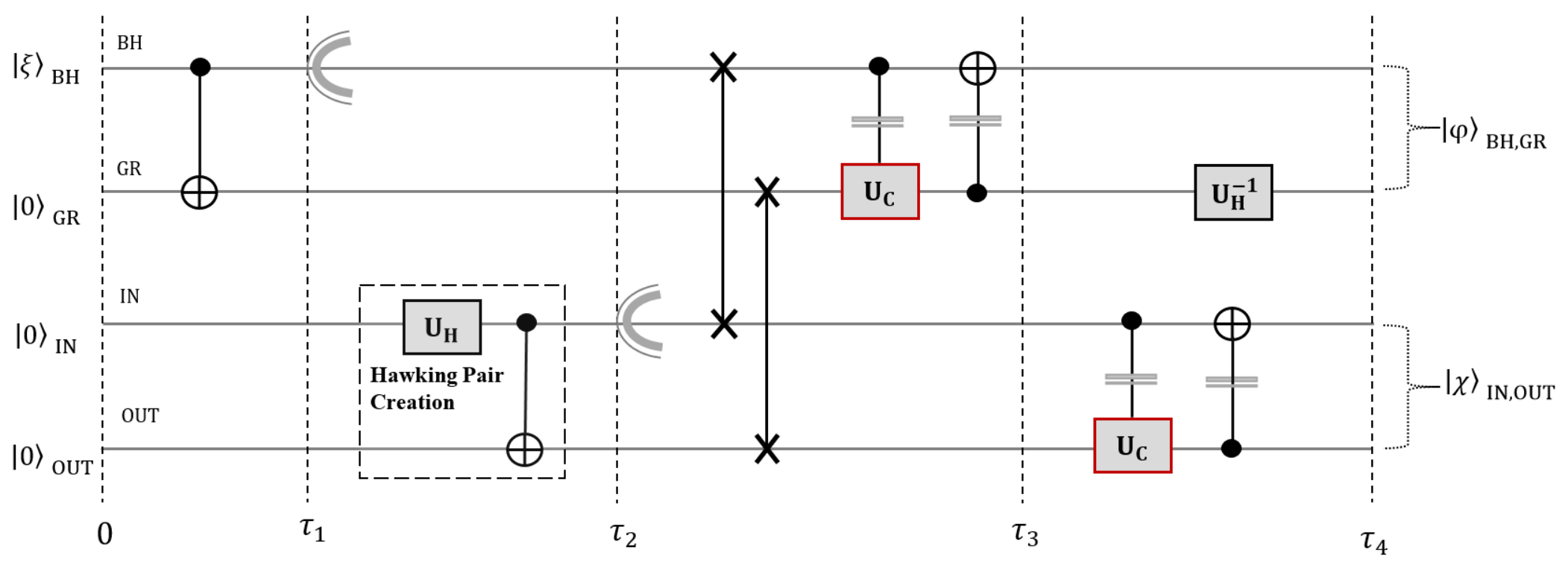}
\caption{Quantum circuit for black hole evaporation with tunable semi-causality violation. The event horizon is indicated by a double grey line with thicker line corresponding to the interior of the black hole. The parametric controlled-unitary gates $\mathrm{CU}(\sigma)$ introduce controlled leakage across the horizon and within the Hawking pair.}
\label{fig:circuit_diagram}
\end{figure}

The circuit evolves over five discrete time steps $\tau_0$ to $\tau_4$, corresponding to: (i) initial state preparation, (ii) pair creation and entanglement generation, (iii) semi-causal evolution, (iv) semi-causality violation via  $\mathrm{CU}(\sigma)$, and (v) final readout or analysis. In the circuit representation, the event horizon is indicated by the double grey line in Fig.~\ref{fig:circuit_diagram}, with the thicker line denoting the black hole interior. The semi-causality condition becomes relevant only after a pair of qubits is separated by the horizon. Consequently, the first two CNOT gates do not violate semi-causality: before their action, the corresponding qubit pairs (BH,GR) and (IN,OUT) are both located outside the horizon. The horizon-crossing interpretation arises only after these gates create the corresponding entangled pairs, with BH and IN subsequently moving into the interior while GR and OUT remain outside.

The full circuit diagram is shown in Fig.~\ref{fig:circuit_diagram}. Our circuit coincides with Broda's circuit~\cite{Broda:2021gts}, except at $\tau_3$ and $\tau_4$ we have inserted two semi-causality violating $\mathrm{CU}(\sigma)$ gates, which is the novelty of our construction. Although the model contains only four qubits, its simplicity permits a fully analytic treatment of the evaporation process and allows the role of the leakage parameter $\sigma$ to be isolated and studied exactly.

\subsection{Implementation of semi-causality violation}
\label{subsec:semi_causality_violation}

To interpolate between the strictly semi-causal limit and a regime with controlled information leakage, we introduce the parametric two-qubit controlled-unitary gate:
\begin{equation}
\mathrm{CU}(\sigma) =
\begin{bmatrix}
1 & 0 & 0 & 0 \\
0 & 1 & 0 & 0 \\
0 & 0 & \cos\sigma & -\sin\sigma \\
0 & 0 & \sin\sigma & \cos\sigma
\end{bmatrix},
\label{eq:cu_gate_def}
\end{equation}
where $\sigma \in [0,\frac{\pi}{4}]$ controls the strength of the semi-causality violation. This matrix is unitary for all $\sigma$. The matrix representation is written in the computational basis
$\{|00\rangle,|01\rangle,|10\rangle,|11\rangle\}$.

Its action on basis states is given by:
\begin{align}
\mathrm{CU}_{A,B}(\sigma) \ket{0}_A\ket{0}_B &= \ket{0}_A\ket{0}_B, \nonumber \\
\mathrm{CU}_{A,B}(\sigma) \ket{0}_A\ket{1}_B &= \ket{0}_A\ket{1}_B, \nonumber \\
\mathrm{CU}_{A,B}(\sigma) \ket{1}_A\ket{0}_B &= \ket{1}_A \otimes \big( \cos\sigma \ket{0} + \sin\sigma \ket{1} \big)_B, \nonumber \\
\mathrm{CU}_{A,B}(\sigma) \ket{1}_A\ket{1}_B &= \ket{1}_A \otimes \big(-\sin\sigma\ket{0} +  \cos\sigma \ket{1} \big)_B.
\label{eq:gate_action}
\end{align}
The last two equations can be rewritten as follows. 
\begin{align}
\mathrm{CU}_{A,B}(\sigma) \ket{1}_A\ket{0}_B &= \cos\sigma  \big( \ket{1}_A \ket{0}_B + \tan\sigma \ket{1}_A \ket{1}_B\big), \nonumber \\
\mathrm{CU}_{A,B}(\sigma) \ket{1}_A\ket{1}_B &= \cos\sigma  \big( \ket{1}_A \ket{1}_B - \tan\sigma \ket{1}_A \ket{0}_B\big).
\label{eq:gate_action2}
\end{align}
The second term in Eq.~(\ref{eq:gate_action2}) represents the component generated by the controlled rotation when the control qubit is in the state $\ket{1}$. When the control qubit is located inside the black hole and the target qubit outside the horizon, the state of the interior subsystem influences the evolution of exterior degrees of freedom. In this sense, the gate introduces a controlled departure from the semi-causal structure of Broda's model, with the strength of the effect governed by $\tan\sigma$.

\begin{itemize}
\item $\sigma = 0$: the gate reduces to identity, enforcing strict semi-causality (no leakage). The circuit reduces to Broda's circuit.
\item $0 < \sigma < \frac{\pi}{4}$: partial leakage of quantum information across the horizon.
\item $\sigma = \frac{\pi}{4}$: maximum semi-causality violation.
\end{itemize}

The interval $\sigma\in[0,\pi/4]$ is sufficient to capture the physically distinct regimes of the gate, with $\sigma=\pi/4$ corresponding to the largest leakage amplitude permitted by the present construction.

This provides a continuous deformation away from Broda's strictly semi-causal circuit and offers a simple framework for exploring the information-theoretic consequences of controlled information leakage across the horizon.

Finally, the first unitary acting on the BH-GR subsystem is
\begin{equation}
U_H =
\begin{bmatrix}
\theta & -\eta \\
\eta & \theta
\end{bmatrix},
\qquad |\theta|^2 + |\eta|^2 = 1,
\end{equation}
which controls the fraction of initial BH information that mixes with near-horizon gravitational modes. The parameters $\theta,\eta\in \mathbb{C}$ may be interpreted as encoding properties such as black hole mass or temperature; for example, maximal mixing occurs at $|\eta| \simeq 1/\sqrt{2}$.

\section{Evaporation dynamics and entanglement structure}
\label{section3}

We now study the dynamical behaviour of quantum correlations in the circuit introduced in Section~2. Our goal is to understand how controlled violations of semi-causality modify entanglement and information flow during black hole evaporation. In the limit $\sigma=0$, our construction reduces exactly to Broda's semi-causal model~\cite{Broda:2021gts}, reproducing its Page-like entropy evolution. The new feature is the emergence of residual entropy and persistent entanglement for $\sigma>0$, which provide quantitative measures of controlled information leakage across the horizon.

\subsection{Evolution of the state}
\label{subsec:evolution}

We now track the evolution of the total quantum state through the sequence of operations shown in Fig.~\ref{fig:circuit_diagram}. At each time step $\tau_i$, we explicitly compute the updated state and discuss the corresponding physical process.

Throughout, we consider the four-qubit register 
\begin{align}
&\mathcal{H}_{\mathrm{BH}} \otimes 
\mathcal{H}_{\mathrm{GR}} \otimes 
\mathcal{H}_{\mathrm{IN}} \otimes 
\mathcal{H}_{\mathrm{OUT}},
\end{align}
initialized in the pure product state 
\begin{align}
&|\Psi(0)\rangle = (\alpha|0\rangle + \beta|1\rangle)_{\mathrm{BH}}
\otimes |0\rangle_{\mathrm{GR}}
\otimes |0\rangle_{\mathrm{IN}}
\otimes |0\rangle_{\mathrm{OUT}},
\qquad |\alpha|^2 + |\beta|^2 = 1.
\end{align}

\paragraph{Time step $\boldsymbol{\tau_1}$:}
A CNOT gate with BH as control and GR as target entangles the interior with the near-horizon gravitational mode:
\begin{align}
|\Psi(\tau_1)\rangle_{\mathrm{BH,GR,IN,OUT}} 
&= \text{CNOT}_{\mathrm{BH,GR}} \left( |\psi\rangle_{\mathrm{BH}} |0\rangle_{\mathrm{GR}} |0\rangle_{\mathrm{IN}} |0\rangle_{\mathrm{OUT}} \right) \nonumber \\
&= \left( \alpha |0\rangle_{\mathrm{BH}} |0\rangle_{\mathrm{GR}} + \beta |1\rangle_{\mathrm{BH}} |1\rangle_{\mathrm{GR}} \right) |0\rangle_{\mathrm{IN}} |0\rangle_{\mathrm{OUT}}.
\label{eq:state_tau1}
\end{align}
This step creates a Bell-like correlation across the horizon, modelling the gravitational dressing of the black hole microstate.

\paragraph{Time step $\boldsymbol{\tau_2}$: }
The unitary $U_H$ acts on the IN qubit, mixing the infalling vacuum into a superposed pair, after which a CNOT (IN$\rightarrow$OUT) produces the Hawking pair:
\begin{align}
|\Psi(\tau_2)\rangle_{\mathrm{BH,GR,IN,OUT}} 
&= \left( \alpha |0\rangle_{\mathrm{BH}} |0\rangle_{\mathrm{GR}} + \beta |1\rangle_{\mathrm{BH}} |1\rangle_{\mathrm{GR}} \right) \otimes \text{CNOT}_{\mathrm{IN,OUT}} \left( U_H |0\rangle_{\mathrm{IN}} \otimes |0\rangle_{\mathrm{OUT}} \right) \nonumber \\
&= \left( \alpha |0\rangle_{\mathrm{BH}} |0\rangle_{\mathrm{GR}} + \beta |1\rangle_{\mathrm{BH}} |1\rangle_{\mathrm{GR}} \right) \otimes \left( \theta |0\rangle_{\mathrm{IN}} |0\rangle_{\mathrm{OUT}} + \eta |1\rangle_{\mathrm{IN}} |1\rangle_{\mathrm{OUT}} \right),
\label{eq:state_tau2}
\end{align}
where $U_H|0\rangle_{\mathrm{IN}} = \theta|0\rangle_{\mathrm{IN}} + \eta|1\rangle_{\mathrm{IN}}$ with $|\theta|^2+|\eta|^2 = 1$.  
This step models Hawking pair creation and the spread of entanglement into the radiation sector. Up to  $\tau_2$, the evolution is identical to Broda's circuit. Deviations first appear at $\tau_3$, where the controlled-unitary gates introduce semi-causality violation.
\paragraph{Time step $\boldsymbol{\tau_3}$:}
Two SWAP gates exchange the pairs BH$\leftrightarrow$IN and GR$\leftrightarrow$OUT, mimicking the infall of the interior mode and the outward propagation of Hawking radiation:
\begin{align}
|\widetilde{\Psi}(\tau_3) \rangle 
= \left( \theta |0\rangle_{\mathrm{BH}} |0\rangle_{\mathrm{GR}} + \eta |1\rangle_{\mathrm{BH}} |1\rangle_{\mathrm{GR}} \right) 
\otimes \left( \alpha |0\rangle_{\mathrm{IN}} |0\rangle_{\mathrm{OUT}} + \beta |1\rangle_{\mathrm{IN}} |1\rangle_{\mathrm{OUT}} \right).
\label{eq:state_after_swap}
\end{align}

Next, the controlled-unitary $\mathrm{CU}_{\mathrm{BH,GR}}(\sigma)$ acts between BH and GR, introducing tunable semi-causality violation:
\begin{align}
|\widetilde{\phi}(\tau_3)\rangle_{\mathrm{BH,GR}} 
&= \mathrm{CU}_{\mathrm{BH,GR}} (\sigma) \left( \theta |0\rangle_{\mathrm{BH}} |0\rangle_{\mathrm{GR}} + \eta |1\rangle_{\mathrm{BH}} |1\rangle_{\mathrm{GR}} \right) \nonumber \\
&=\theta |0\rangle_{\mathrm{BH}} |0\rangle_{\mathrm{GR}} 
- \eta \sin\sigma |1\rangle_{\mathrm{BH}} |0\rangle_{\mathrm{GR}} 
+ \eta \cos\sigma |1\rangle_{\mathrm{BH}} |1\rangle_{\mathrm{GR}}\cr
\end{align}
A CNOT$_{\mathrm{GR,BH}}$ then redistributes correlations within the pair:
\begin{align}
|\phi(\tau_3)\rangle_{\mathrm{BH,GR}} 
&=\text{CNOT}_{\mathrm{GR,BH}} |\widetilde{\phi}(\tau_3)\rangle_{\mathrm{BH,GR}} \cr
&=\theta |0\rangle_{\mathrm{BH}} |0\rangle_{\mathrm{GR}} 
- \eta \sin\sigma |1\rangle_{\mathrm{BH}} |0\rangle_{\mathrm{GR}} 
+ \eta \cos\sigma|0\rangle_{\mathrm{BH}} |1\rangle_{\mathrm{GR}}.
\label{eq:state_bhgr_tau3}
\end{align}

Thus, the total state at $\tau_3$ is:
\begin{align}
|\Psi(\tau_3)\rangle_{\mathrm{BH,GR,IN,OUT}} 
&= |\phi(\tau_3)\rangle_{\mathrm{BH,GR}} \otimes \left( \alpha |0\rangle_{\mathrm{IN}} |0\rangle_{\mathrm{OUT}} + \beta |1\rangle_{\mathrm{IN}} |1\rangle_{\mathrm{OUT}} \right).
\label{eq:state_tau3}
\end{align}

\paragraph{Time step $\boldsymbol{\tau_4}$: }
The inverse unitary $U_H^{-1}$ acts on GR, partially undoing the earlier infalling mixing:
\begin{align}
|\phi(\tau_4)\rangle_{\mathrm{BH,GR}} &= U_H^{-1} |\phi(\tau_3)\rangle_{\mathrm{BH,GR}} \nonumber \\
&= (|\theta|^2+|\eta|^2\cos\sigma) |0\rangle_{\mathrm{BH}} |0\rangle_{\mathrm{GR}} 
+(\theta^* \eta\cos\sigma-\theta \eta^*)|0\rangle_{\mathrm{BH}} |1\rangle_{\mathrm{GR}} 
\nonumber \\
&\quad - \theta^*\eta\sin\sigma |1\rangle_{\mathrm{BH}} |0\rangle_{\mathrm{GR}}  +|\eta|^2\sin\sigma |1\rangle_{\mathrm{BH}} |1\rangle_{\mathrm{GR}}.
\label{eq:state_bhgr_tau4}
\end{align}
Simultaneously, the same controlled-unitary $\mathrm{CU}(\sigma)$ gate acts on IN-OUT, followed by a CNOT$_{\mathrm{OUT,IN}}$:
 \begin{align}
 |\chi\rangle_{\mathrm{IN,OUT}}(\tau_4)
 &=\alpha|0\rangle_{\mathrm{IN}}|0\rangle_{\mathrm{OUT}}+\beta\cos\sigma|0\rangle_{\mathrm{IN}}|1\rangle_{\mathrm{OUT}}-\beta\sin\sigma|1\rangle_{\mathrm{IN}}|0\rangle_{\mathrm{OUT}}
 \end{align}
 The final four-qubit state is therefore
\begin{align}
|\Psi(\tau_4)\rangle_{\mathrm{BH,GR,IN,OUT}} 
= |\phi(\tau_4)\rangle_{\mathrm{BH,GR}} \otimes |\chi(\tau_4)\rangle_{\mathrm{IN,OUT}}.
\label{eq:state_tau4}
\end{align}
A useful consistency check is obtained in the limit $\sigma=0$, where all controlled-unitary gates reduce to the identity. In this limit our circuit becomes identical to Broda's original semi-causal evaporation model~\cite{Broda:2021gts}, and the final state reduces to:
\begin{align}
|\Psi(\tau_4)\rangle_{\mathrm{BH,GR,IN,OUT}}|_{\sigma\rightarrow 0}
&=\ket{0}_{\mathrm{BH}}
  \otimes  \ket{0}_{\mathrm{GR}}
  \otimes \ket{0}_{\mathrm{IN}}
  \otimes (\alpha \ket{0} + \beta \ket{1})_{\mathrm{OUT}}\cr
  &=\ket{0}_{\mathrm{BH}}
  \otimes  \ket{0}_{\mathrm{GR}}
  \otimes \ket{0}_{\mathrm{IN}}
  \otimes \ket{\xi}_{\mathrm{OUT}},
\end{align}
showing complete information transfer into the outgoing Hawking radiation. 

We observe that throughout the entire evolution, the total four-qubit state factorizes as 
\begin{align}
&|\Psi(\tau_i)\rangle_{\mathrm{BH,GR,IN,OUT}} = |\phi(\tau_i)\rangle_{\mathrm{BH,GR}} \otimes |\chi(\tau_i)\rangle_{\mathrm{IN,OUT}},
\end{align}
Consequently, all entanglement measures considered below arise from
correlations within the BH--GR and IN--OUT sectors, with no contribution
from cross-sector entanglement. This factorization is a direct consequence
of the circuit architecture and remains valid even in the presence of
the semi-causality-violating gates.

\subsection{Entanglement dynamics}
\label{subsec:entanglement-dynamics}

To understand how quantum correlations evolve during black hole evaporation in our circuit model, we evaluate three standard entanglement measures at each time step:

\begin{itemize}
    \item \textbf{von Neumann entropy} of a reduced density matrix $\rho$, quantifying its mixedness:
    \begin{equation}
    S(\rho) = -\Tr(\rho \log \rho),
    \label{eq:entropy_def}
    \end{equation}

    \item \textbf{Mutual information} between subsystems $A$ and $B$, capturing the total (classical and quantum) correlations:
    \begin{equation}
    \mathcal{I}(A:B) = S_A + S_B - S_{AB},
    \label{eq:mutual_info_def}
    \end{equation}

    \item \textbf{Entanglement negativity}~\cite{Vidal:2002zz,He_2015,Tokusumi:2018typ}, which detects quantum entanglement between mixed-state subsystems:
    \begin{equation}
    \mathcal{N}(A:B) = \frac{ \lVert \rho_{AB}^{T_B} \rVert_1 - 1 }{2}
    = \frac{1}{2} \left( \sum_i |\lambda_i^{T}| - 1 \right),
    \label{eq:negativity_def}
    \end{equation}
    where $\rho_{AB}^{T_B}$ is the partial transpose of $\rho_{AB}$ with respect to subsystem $B$, and $\lambda_i^T$ are its eigenvalues.
\end{itemize}

Entanglement negativity is particularly convenient in the present setting because it provides a simple measure of bipartite entanglement that remains applicable to both pure and mixed states. Since the reduced density matrices encountered during the evaporation process are generally mixed, negativity offers a uniform characterization of entanglement throughout the circuit evolution.

We apply these measures to two natural bi-partitions: 
(i) the near-horizon pair $(\mathrm{BH}, \mathrm{GR})$ and 
(ii) the emitted Hawking pair $(\mathrm{IN}, \mathrm{OUT})$.

The parameter $\sigma$ controls the strength of semi-causality violation introduced via the controlled-unitary gates. When $\sigma = 0$, the circuit respects standard causal structure. Increasing $\sigma$ leads to stronger backward information flow, modifying entanglement patterns.

We now summarize the analytic expressions for the von Neumann entropy, mutual information, and entanglement negativity at each time step. Since the controlled-unitary gates first appear at $\tau_3$, the results at $\tau_1$ and $\tau_2$ coincide with those of Broda's semi-causal circuit~\cite{Broda:2021gts}. The new features of our model emerge at $\tau_3$ and $\tau_4$, where the semi-causality-violating parameter $\sigma$ modifies the entanglement structure and generates persistent late-time correlations. Full derivations are presented in Appendix~\ref{appendix:a}.

\paragraph{Time $\tau_1$:} A CNOT gate creates entanglement between BH and GR:
\begin{align}
S_{\mathrm{BH}}(\tau_1) &= -|\alpha|^2 \log |\alpha|^2 - |\beta|^2 \log |\beta|^2,
\label{eq:s_bh_tau1} \\
S_{\mathrm{OUT}}(\tau_1) &= 0,
\label{eq:s_out_tau1} \\
\mathcal{I}_{\mathrm{BH:GR}}(\tau_1) &= 2 S_{\mathrm{BH}}(\tau_1),
\label{eq:mi_bhgr_tau1} \\
\mathcal{N}_{\mathrm{BH:GR}}(\tau_1) &= |\alpha||\beta|, \\
\mathcal{I}_{\mathrm{IN:OUT}}(\tau_1) &= 0, \quad 
\mathcal{N}_{\mathrm{IN:OUT}}(\tau_1) = 0.
\end{align}

\paragraph{Time $\tau_2$:} A unitary $U_H$ followed by a CNOT between IN and OUT creates entanglement in the radiation subsystem:
\begin{align}
S_{\mathrm{BH}}(\tau_2) &= S_{\mathrm{BH}}(\tau_1),
\label{eq:s_bh_tau2} \\
S_{\mathrm{OUT}}(\tau_2) &= -|\theta|^2 \log |\theta|^2 - |\eta|^2 \log |\eta|^2,
\label{eq:s_out_tau2} \\
\mathcal{I}_{\mathrm{BH:GR}}(\tau_2) &= \mathcal{I}_{\mathrm{BH:GR}}(\tau_1), \\
\mathcal{N}_{\mathrm{BH:GR}}(\tau_2) &= \mathcal{N}_{\mathrm{BH:GR}}(\tau_1), \\
\mathcal{I}_{\mathrm{IN:OUT}}(\tau_2) &= 2 S_{\mathrm{OUT}}(\tau_2),
\label{eq:mi_inout_tau2} \\
\mathcal{N}_{\mathrm{IN:OUT}}(\tau_2) &= |\theta||\eta|.
\label{eq:neg_inout_tau2}
\end{align}
Up to this stage, the circuit evolution is identical to that of Broda's model. The BH-GR and IN-OUT subsystems form two independent entangled pairs, and no effects associated with semi-causality violation are present.

\paragraph{Time $\tau_3$:} The controlled-unitary $\mathrm{CU}(\sigma)$ acting on BH-GR modifies their correlations:
\begin{align}
S_{\mathrm{BH}}(\tau_3) &= -\log(|\eta|^2 \cos\sigma \sin\sigma) \cr
&\quad - \frac{1}{2} \sqrt{1 - 4 |\eta|^4 \cos^2\sigma \sin^2\sigma} \log \left( \frac{1 + \sqrt{1 - 4 |\eta|^4 \cos^2\sigma \sin^2\sigma}}{1 - \sqrt{1 - 4 |\eta|^4\cos^2\sigma \sin^2\sigma}} \right),
\label{eq:s_bh_tau3} \\
S_{\mathrm{OUT}}(\tau_3) &= -|\alpha|^2 \log |\alpha|^2 - |\beta|^2 \log |\beta|^2,
\label{eq:s_out_tau3} \\
\mathcal{I}_{\mathrm{BH:GR}}(\tau_3) &= 2 S_{\mathrm{BH}}(\tau_3),
\label{eq:mi_bhgr_tau3} \\
\mathcal{I}_{\mathrm{IN:OUT}}(\tau_3) &= 2 S_{\mathrm{OUT}}(\tau_3),
\label{eq:mi_inout_tau3} \\
\mathcal{N}_{\mathrm{BH:GR}}(\tau_3) &= |\eta|^2 \cos\sigma \sin\sigma,
\label{eq:neg_bhgr_tau3} \\
\mathcal{N}_{\mathrm{IN:OUT}}(\tau_3) &= |\alpha||\beta|.
\label{eq:neg_inout_tau3}
\end{align}

Unlike the semi-causal case, the controlled-unitary gate introduces an explicit dependence on $\sigma$ into the BH-GR correlations. As a result, the entropy and negativity no longer vanish after information transfer, signaling the onset of persistent correlations associated with controlled information leakage.

\paragraph{Time $\tau_4$:} The second controlled-unitary and CNOT gate modify IN-OUT entanglement. BH-GR values remain unchanged:
\begin{align}
S_{\mathrm{BH}}(\tau_4) &= S_{\mathrm{BH}}(\tau_3), \quad
\mathcal{I}_{\mathrm{BH:GR}}(\tau_4) = \mathcal{I}_{\mathrm{BH:GR}}(\tau_3), \quad
\mathcal{N}_{\mathrm{BH:GR}}(\tau_4) = \mathcal{N}_{\mathrm{BH:GR}}(\tau_3),
\label{eq:bhgr_tau4_same}
\end{align}
\begin{align}
S_{\mathrm{OUT}}(\tau_4) 
&= -\log(|\beta|^2 \cos\sigma \sin\sigma) \cr
&\quad - \frac{1}{2} \sqrt{1 - 4 |\beta|^4 \cos^2\sigma \sin^2\sigma} \log \left( \frac{1 + \sqrt{1 - 4 |\beta|^4 \cos^2\sigma \sin^2\sigma}}{1 - \sqrt{1 - 4 |\beta|^4\cos^2\sigma \sin^2\sigma}} \right),
\label{eq:s_out_tau4} \\
\mathcal{I}_{\mathrm{IN:OUT}}(\tau_4) &= 2 S_{\mathrm{OUT}}(\tau_4),
\label{eq:mi_inout_tau4} \\
\mathcal{N}_{\mathrm{IN:OUT}}(\tau_4) &= |\beta|^2 \cos\sigma \sin\sigma.
\label{eq:neg_inout_tau4}
\end{align}

For $\sigma=0$, the late-time entanglement measures vanish and the circuit reproduces Broda's semi-causal information transfer, yielding complete purification of the outgoing radiation. In contrast, for any nonzero value of $\sigma$, the reduced BH and OUT subsystems retain finite entropy, while the corresponding BH-GR and IN-OUT correlations remain nonzero at late times. These residual correlations constitute the central new feature of the present model and provide a quantitative measure of semi-causality violation.

\subsection{Page-like behaviour and residual entropy}
\label{subsec:numerical-results}

To understand the effect of semi-causality violation on the evaporation
process, we now examine the entropy evolution of the black hole and
radiation subsystems. In the limit $\sigma=0$, our circuit reduces exactly
to Broda's semi-causal model~\cite{Broda:2021gts}, reproducing its
Page-like behaviour in which the entropy rises during evaporation and
returns to zero at the final time step. Similar behaviour has also been
observed in other quantum circuit models of black hole evaporation
~\cite{Tokusumi:2018typ}, where complete information recovery is reflected
in the vanishing of the late-time entropy. The central new feature of our
model is that for any $\sigma>0$, the entropy no longer returns to zero but
instead saturates at a finite value, indicating persistent late-time
correlations.

We define the diagnostic quantity
\begin{align}
S_{\mathrm{BH+OUT}}(\tau)
\equiv
S_{\mathrm{BH}}(\tau)
+
S_{\mathrm{OUT}}(\tau),
\label{eq:SBHOUT}
\end{align}
where $S_{\mathrm{BH}}$ and $S_{\mathrm{OUT}}$ denote the von Neumann
entropies of the reduced BH and OUT qubits, respectively. The first term
characterizes the entanglement between the black hole interior and the
near-horizon gravitational degrees of freedom, while the second
characterizes the entanglement between the outgoing Hawking radiation
and its infalling partner. Since the total four-qubit state factorizes
into the BH-GR and IN-OUT sectors, both two-qubit sectors are pure at
every time step. Consequently,
\begin{align*}
S_{\mathrm{BH}}=S_{\mathrm{GR}},
\qquad
S_{\mathrm{OUT}}=S_{\mathrm{IN}}.
\end{align*}
Thus, $S_{\mathrm{BH+OUT}}$ is a sum of reduced entropies associated with
the two bipartite sectors, rather than the von Neumann entropy of either
complete two-qubit sector or of the global state.

One may alternatively consider $S_{\mathrm{OUT}}$ alone as the entropy
associated with the Hawking radiation and use it to characterize the
Page-curve behaviour. Since the IN-OUT sector is a pure bipartite state,
$S_{\mathrm{OUT}}$ measures the entanglement between the outgoing
radiation and its infalling partner. This choice is closer to the
original Page-curve interpretation and leads, in the present model, to
the same qualitative conclusions regarding the residual late-time
correlations generated by semi-causality violation.

\begin{figure}[t]
\centering
\includegraphics[width=0.70\textwidth]{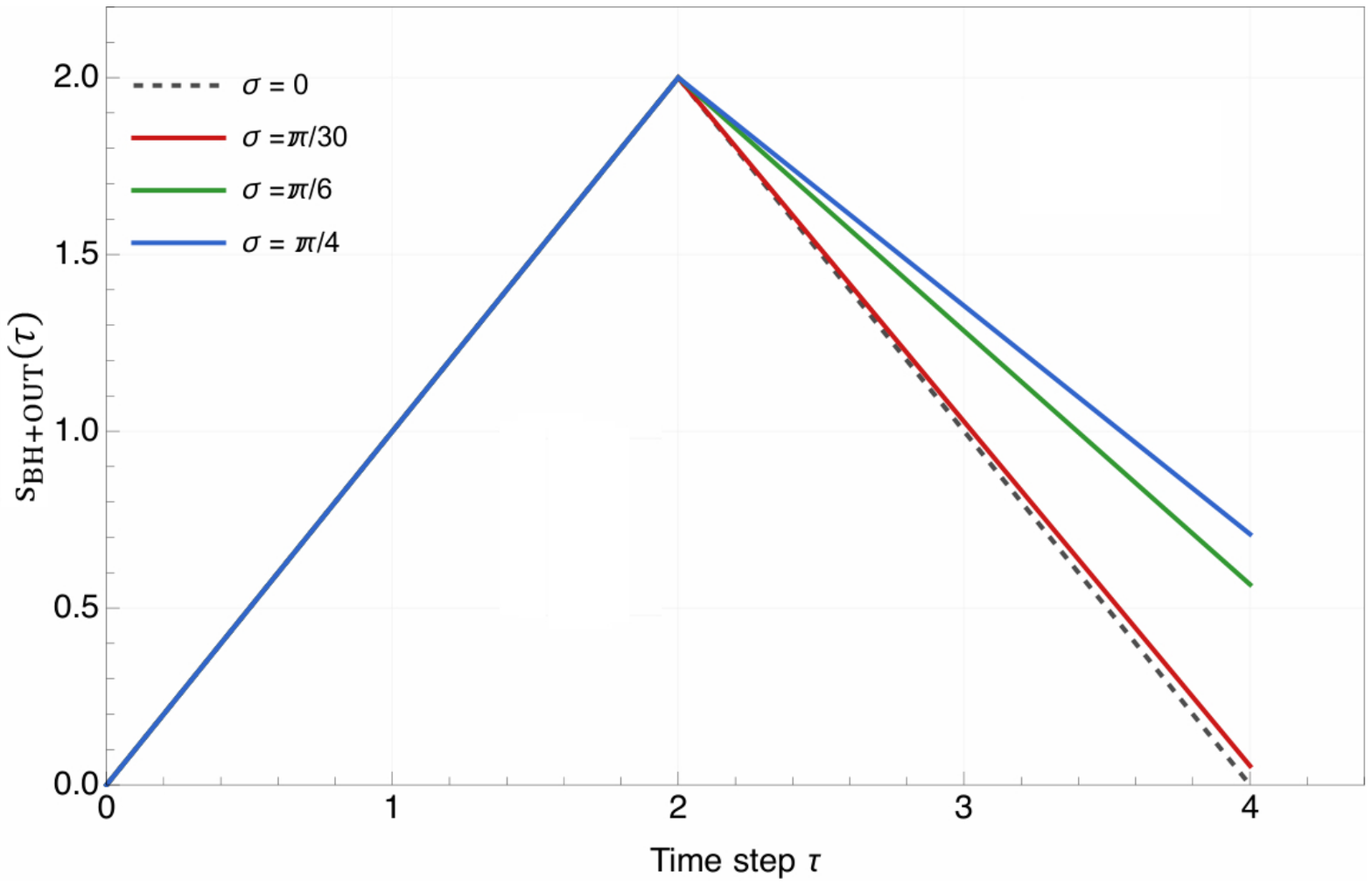}
\caption{Sum of reduced entropies
$S_{\mathrm{BH+OUT}}=S_{\mathrm{BH}}+S_{\mathrm{OUT}}$ defined in
Eq.~\eqref{eq:SBHOUT} as a function of the discrete time step $\tau$ for
different values of the semi-causality violation parameter $\sigma$.
For $\sigma=0$, $S_{\mathrm{BH+OUT}}$ returns to zero at late times,
whereas for $\sigma>0$ a nonzero residual value remains.}
\label{fig:entropy_pagecurve}
\end{figure}

To generate the plots in this subsection, we set the parameters $|\theta| = |\eta| = 1/\sqrt{2}$ and $|\alpha| = |\beta| = 1/\sqrt{2}$, corresponding to maximally entangled initial and intermediate states. We have verified that the qualitative features, such as entropy saturation and persistent correlations, remain robust across different parameter choices.

In Figure~\ref{fig:entropy_pagecurve}, we show the sum of the reduced
entropies $S_{\mathrm{BH+OUT}}$ as a function of time for several values
of $\sigma$. When $\sigma=0$, the entropy follows a clean Page-like curve,
peaking at the Page time and returning to zero at $\tau=4$. For
$\sigma>0$, the final value of $S_{\mathrm{BH+OUT}}$ remains nonzero even
though the overall evolution is unitary. In contrast to Broda's model and
other Page-curve constructions, complete purification of the outgoing
radiation does not occur. Instead, the circuit retains finite late-time
entanglement, whose magnitude increases with the strength of the
semi-causality violation.

\begin{figure}[t]
\centering

\begin{subfigure}{0.48\textwidth}
\includegraphics[width=\linewidth]{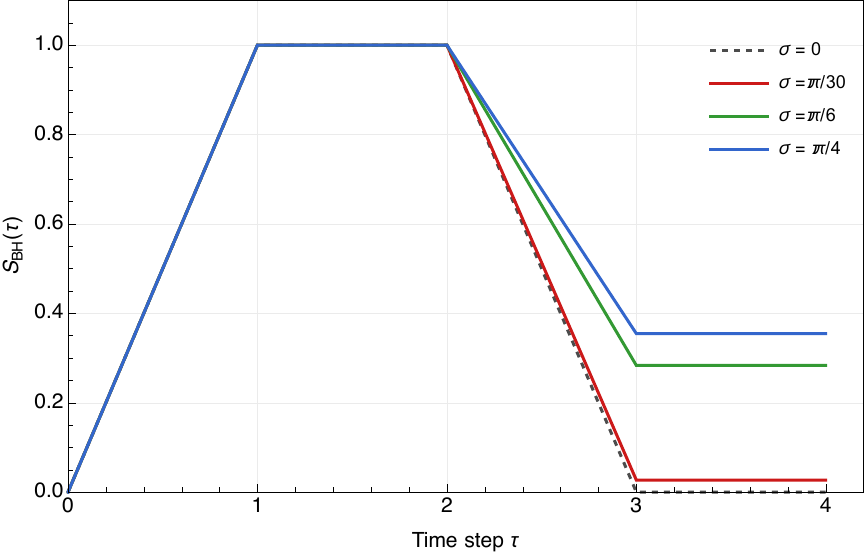}
\caption{$S_{\mathrm{BH}}(\tau)$}
\end{subfigure}
\hfill
\begin{subfigure}{0.48\textwidth}
\includegraphics[width=\linewidth]{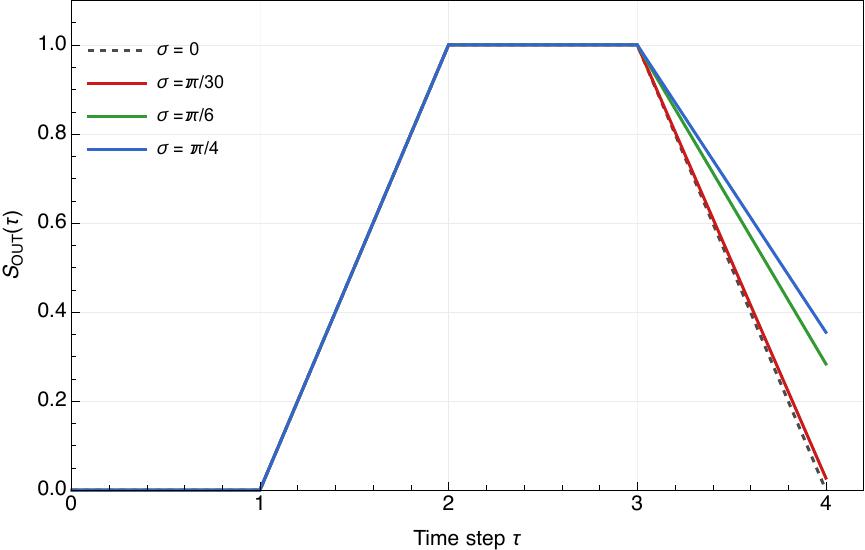}
\caption{$S_{\mathrm{OUT}}(\tau)$}
\end{subfigure}

\vspace{0.5em}

\begin{subfigure}{0.48\textwidth}
\includegraphics[width=\linewidth]{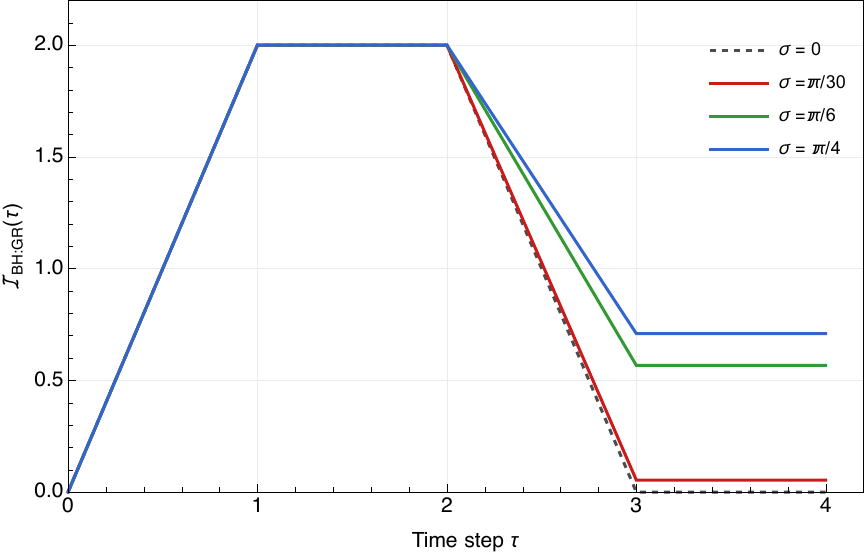}
\caption{$\mathcal{I}_{\mathrm{BH:GR}}(\tau)$}
\end{subfigure}
\hfill
\begin{subfigure}{0.48\textwidth}
\includegraphics[width=\linewidth]{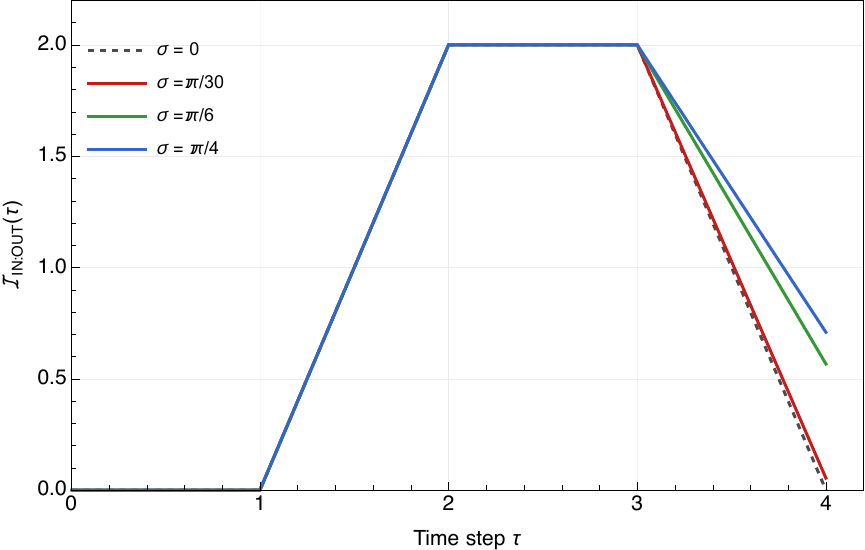}
\caption{$\mathcal{I}_{\mathrm{IN:OUT}}(\tau)$}
\end{subfigure}

\vspace{0.5em}

\begin{subfigure}{0.48\textwidth}
\includegraphics[width=\linewidth]{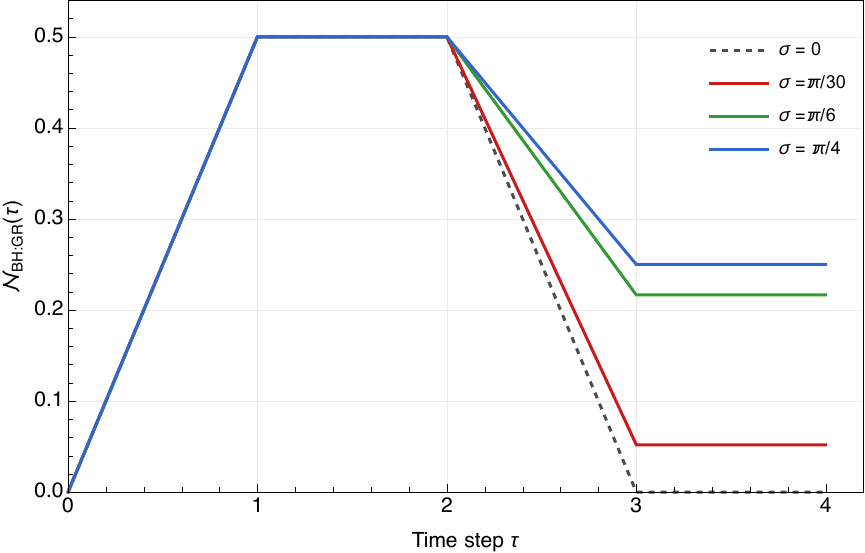}
\caption{$\mathcal{N}_{\mathrm{BH:GR}}(\tau)$}
\end{subfigure}
\hfill
\begin{subfigure}{0.48\textwidth}
\includegraphics[width=\linewidth]{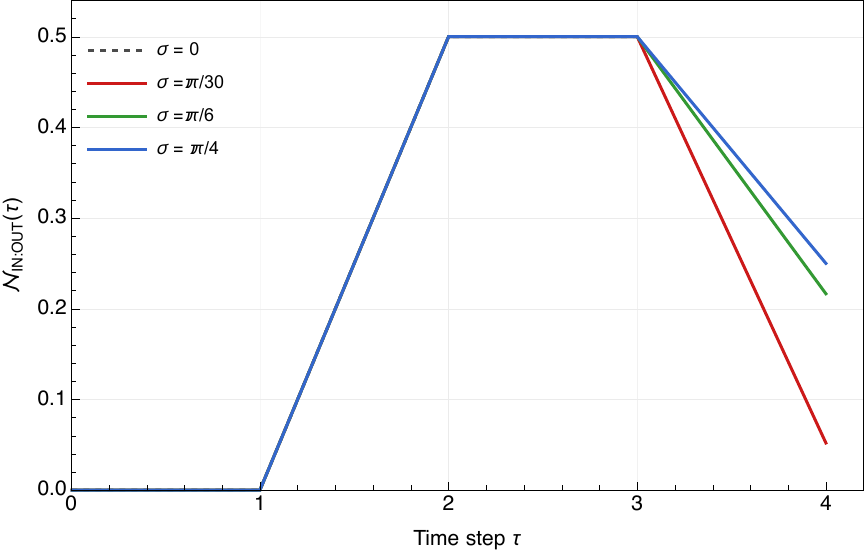}
\caption{$\mathcal{N}_{\mathrm{IN:OUT}}(\tau)$}
\end{subfigure}

\caption{
Entanglement measures evaluated at time steps $\tau$ for different values of
$\sigma$. Rows show (from top to bottom) von Neumann entropy,
mutual information, and entanglement negativity. Columns correspond to the
two bipartitions BH
-GR and IN-OUT.}
\label{fig:combined_entanglement}
\end{figure}

To understand the entanglement structure in more detail, we analyze bipartite correlations between BH-GR and IN-OUT across time. Figure~\ref{fig:combined_entanglement} shows the evolution of von Neumann entropy, mutual information, and entanglement negativity for both bi-partitions. Each row corresponds to a different entanglement measure, while the columns separate the two bi-partitions. 

The persistence of both mutual information and negativity at $\tau_4$ is a distinctive consequence of the controlled-unitary deformation introduced in our model. In the semi-causal limit ($\sigma=0$), these quantities vanish at late times, reproducing the behaviour of Broda's original circuit~\cite{Broda:2021gts}. For $\sigma>0$, however, both mutual information and negativity remain finite, providing direct measures of residual information retention and incomplete purification of the emitted radiation.

\begin{figure}[t]
\centering
\includegraphics[width=0.7\textwidth]{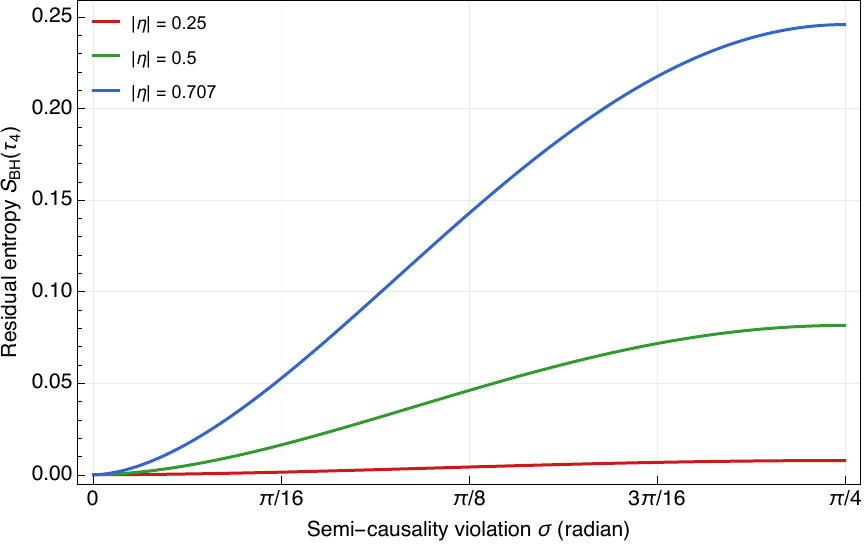}
\caption{
Residual entropy $S_{\mathrm{BH}}(\tau_4)$ as a function of the semi-causality
violation parameter $\sigma$ for several values of the initial
entanglement parameter $|\eta|$. Curves are shown for
$|\eta| \in \{0.25,\,0.5,\,1/\sqrt{2}\}$.}
\label{fig:residual_entropy_sigma}
\end{figure}

Finally, in Figure~\ref{fig:residual_entropy_sigma}, we focus on the late-time residual entropy $S_{\mathrm{BH}}(\tau_4)$ as a function of the semi-causality violation parameter $\sigma$, for three different values of the initial entanglement parameter $\eta$. We find that the residual entropy increases monotonically with $\sigma$, indicating that stronger semi-causality violation leads to greater late-time information retention. Lower values of $|\eta|$ yield smaller residual entropy, as expected from the reduced initial entanglement between BH and GR.

These results show that controlled departures from the semi-causal limit generate a finite residual entropy whose magnitude is governed by the parameter $\sigma$. This residual structure provides the basis for the heuristic connections to modified black hole evaporation scenarios discussed in Section~\ref{section4}.

It is worth emphasizing that the residual entropy observed here is not a consequence of nonunitary evolution. The entire circuit remains unitary for all values of $\sigma$. Rather, the entropy plateau arises because controlled semi-causality violation prevents complete transfer of quantum correlations into the outgoing radiation sector.

\section{Connections to black hole evaporation scenarios}
\label{section4}

We now interpret our quantum circuit model from a space-time perspective. Having shown that semi-causality violation ($\sigma > 0$) leads to persistent late-time entropy and entanglement, it is natural to ask whether these features have qualitative counterparts in existing scenarios of black hole evaporation.

Our goal in this section is to connect the behaviour of the circuit with several scenarios of black hole evaporation that exhibit modified late-time dynamics and information retention. The resulting entropy plateau and persistent correlations exhibit qualitative similarities to logarithmic entropy corrections and remnant-like information retention that have been discussed in various models of modified black hole evaporation. We use the circuit as a simple information-theoretic framework whose entanglement structure exhibits heuristic similarities with such scenarios. From this perspective, the parameter $\sigma$ may be viewed as a measure of the departure from the strictly semi-causal limit.

\subsection{Logarithmic entropy corrections and remnant-like behaviour}

A common feature of many proposed modifications of black hole evaporation is the appearance of corrections to the Bekenstein--Hawking entropy.

In particular, analyses based on the Generalized Uncertainty Principle (GUP) ~\cite{Maggiore:1993rv,Scardigli:1999jh,Veneziano:1986zf,Adler:2001vs,Chen:2014jwq} yield a modified entropy-area relation of the form   
\begin{align}
S_{\rm GUP} \;=\; \frac{A}{4\,L_p^2}
  \;+\; a_0 \ln\!\left(\frac{A}{4\,L_p^2}\right)
  \;+\; \sum_{n=1}^{\infty} a_n \left(\frac{A}{4\,L_p^2}\right)^{-n}
  \;+\;\text{constant},
\label{eq:GUPentropy}
\end{align}
where $A$ is the horizon area, and the coefficients $a_0$ and $a_n$ depend on the underlying model
\cite{Adler:2001vs,Scardigli:1999jh,Faizal:2014tea,Medved:2004yu}.
Such logarithmic corrections become increasingly important near the endpoint of evaporation and are frequently associated with remnant-like information retention.

In our circuit model, the late-time black hole entropy is given by
\begin{align}
S_{\mathrm{BH}}(\tau_4)
&=-\log(|\eta|^2 \cos\sigma \sin\sigma) \cr
&- \frac{1}{2} \sqrt{1 - 4 |\eta|^4 \cos^2\sigma \sin^2\sigma} \log \left( \frac{1 + \sqrt{1 - 4 |\eta|^4 \cos^2\sigma \sin^2\sigma}}{1 - \sqrt{1 - 4 |\eta|^4\cos^2\sigma \sin^2\sigma}} \right).
\label{eq:SBH_tau4}
\end{align}
In the small-$\sigma$ limit ($\sigma \ll 1$), this reduces to
\begin{align}
S_{\mathrm{plateau}} \;\sim\; -\,|\eta|^4\,\sigma^2 \ln\!\big(2\,|\eta|^4\,\sigma^2\big)
                   \;+\; |\eta|^4\,\sigma^2
                   \;+\;\mathcal O(\sigma^4).
\label{eq:Splateau_small_sigma}
\end{align}

The logarithmic contribution in Eq.~\eqref{eq:Splateau_small_sigma} bears a qualitative resemblance to the logarithmic corrections appearing in Eq.~\eqref{eq:GUPentropy}. In many GUP-based scenarios, such corrections become increasingly important near the endpoint of evaporation and can lead to the formation of a finite-entropy remnant~\cite{Adler:2001vs,Xiang:2006mg}. Similarly, the small-$\sigma$ regime of our circuit exhibits a nonvanishing entropy plateau, indicating incomplete information transfer at late times. From this perspective, the entropy plateau observed in the circuit may be viewed as a simple information-theoretic analogue of remnant-like information retention.

We note that this correspondence is heuristic rather than quantitative. The circuit does not provide a derivation of GUP-corrected black hole thermodynamics. Rather, the similarity lies in the appearance of logarithmic entropy corrections and the persistence of nonvanishing late-time entropy. Nevertheless, the similarity of the entropy scaling suggests that controlled departures from the strictly semi-causal evaporation channel can provide a useful toy-model setting for exploring mechanisms that obstruct complete late-time purification.

\subsection{Analogy with regular black hole remnants}

Another class of modified black hole evaporation scenarios arises in regular black hole geometries, where quantum gravity effects are incorporated directly into the spacetime metric through the introduction of a nonsingular core~\cite{Hayward:2005gi,Modesto:2004xx,Frolov:2016pav,Bardeen:1968}. A characteristic feature of many such models is the existence of an extremal configuration with finite entropy and vanishing Hawking temperature, which acts as the endpoint of evaporation~\cite{Dong_2025,Chen:2014jwq}.

The late-time behaviour of our circuit exhibits a qualitative resemblance to this picture. For $\sigma>0$, the final black hole entropy $S_{\mathrm{BH}}(\tau_4)$ remains nonzero, indicating that complete information transfer to the radiation sector does not occur. Instead, the evaporation process terminates with a finite residual entropy determined by the parameters $(\eta,\sigma)$.

From this perspective, the entropy plateau observed in the circuit may be viewed as an information-theoretic analogue of the finite entropy associated with extremal regular black hole remnants. In both cases, the endpoint of evaporation is characterized by a nonvanishing entropy rather than complete disappearance of the black hole degrees of freedom.

The residual entropy $S_{\mathrm{BH}}(\tau_4)$ and the nonzero negativity between BH and GR (see Fig.~\ref{fig:combined_entanglement}) serve as circuit-theoretic indicators of persistent information retention. We find that the survival of quantum correlations at late times prevents complete purification of the OUT subsystem, so that the emitted radiation does not encode all the information about the initial black hole state.

We therefore view the $\sigma>0$ regime as a simple information-theoretic analogue of modified black hole evaporation scenarios in which a finite amount of information remains inaccessible at late times. The resulting entropy plateau and persistent entanglement exhibit qualitative similarities to remnant-like behaviour and other nontrivial endpoint dynamics discussed in the black hole information literature.

\section{Summary and outlook}
\label{section5}

In this work we developed a four-qubit quantum circuit model of black hole evaporation that incorporates a tunable violation of semi-causality via an interior-exterior controlled-unitary gate $\mathrm{CU}(\sigma)$. The single parameter $\sigma$ interpolates between strictly semi-causal evolution and maximal departure from the semi-causal limit, providing a minimal setting to study controlled departures from one-way signalling at the horizon. We computed von Neumann entropy, mutual information, and entanglement negativity at each discrete time step and showed that any $\sigma>0$ leads to a nonzero residual entropy and persistent late-time entanglement across the horizon.

The entropy plateau exhibits a characteristic $-\sigma^{2}\ln\sigma^{2}$ scaling in the small-$\sigma$ regime, reminiscent of negative logarithmic corrections from generalized uncertainty principle scenarios. 
For intermediate and large $\sigma$, the persistence of a finite residual entropy exhibits qualitative similarities to regular or extremal black hole scenarios, where evaporation terminates in a finite-entropy configuration.

Our construction is intentionally minimal. By working with a four-qubit circuit, we are able to obtain exact analytic expressions for the entanglement entropy, mutual information, and negativity throughout the evaporation process. This analytic tractability allows us to isolate the effects of the leakage parameter $\sigma$ and to characterize transparently how controlled semi-causality violation modifies the entanglement structure of black hole evaporation. We therefore view the present model as a simple information-theoretic framework that provides analytic insight into the role of controlled semi-causality violation.

At the same time, we emphasize that the model has important limitations. Owing to the circuit architecture, the total state factorizes into BH-GR and IN-OUT sectors throughout the evolution, preventing the development of more general multipartite entanglement structures. As a result, the model captures only a restricted form of information retention and should not be regarded as a complete description of black hole evaporation. Rather, it serves as a minimal framework in which the consequences of controlled semi-causality violation can be studied analytically.

Several extensions are possible. Following the discussion of Broda's original model~\cite{Broda:2021gts}, we may consider tensor products of the present four-qubit building block to construct larger systems with smoother Page curves and a finer-grained evaporation history. More generally, we would like to explore larger quantum circuits with additional qubits and nontrivial cross-entanglement between subsystems, similar in spirit to recent quantum circuit models of black hole evaporation~\cite{Tokusumi:2018typ}. Such extensions could provide a closer approximation to realistic scrambling dynamics while retaining the information-theoretic perspective developed in this work. Finally, because the present model uses only elementary two-qubit gates, it may be directly implementable on current quantum hardware, enabling experimental studies of residual entropy and horizon leakage in controlled laboratory settings.

\section*{Acknowledgments}

We would like to thank Hikaru Kawai and Pisin Chen for insightful discussions.
SB also thanks Suparna Ballav for assistance in preparing the diagrams.
This work is supported in part by Taiwan’s Ministry of Science and Technology
under grant numbers NSTC 112-2112-M-033-003-MY3 and NSTC 113-2112-M-033-003,
and by the National Center for Theoretical Sciences (NCTS).

\appendix
\section{Details of entanglement computation}
\label{appendix:a}
 
In this appendix we present the explicit density matrices, partial traces, and eigenvalue decompositions used to compute the von Neumann entropy, mutual information, and entanglement negativity at each time step of the four-qubit evaporation circuit. The subsystems are denoted as follows: $\mathrm{BH}$ (black hole interior), $\mathrm{GR}$ (gravitational mode), $\mathrm{IN}$ (infalling partner), and $\mathrm{OUT}$ (outgoing radiation). The calculations are presented sequentially at each time step $\tau_1$--$\tau_4$, highlighting how the controlled-unitary deformation modifies the reduced density matrices and entanglement measures. For $\sigma=0$, all expressions presented below reduce to the corresponding results of Broda's semi-causal circuit~\cite{Broda:2021gts}.

\subsection*{ BH-GR entanglement at \boldmath$\tau_1$}

At $\tau_1$, the circuit applies a CNOT gate between BH and GR. The resulting state is:
\begin{align}
|\Psi(\tau_1)\rangle_{\mathrm{BH,GR,IN,OUT}} 
&= \left( \alpha |0\rangle_{\mathrm{BH}} |0\rangle_{\mathrm{GR}} + \beta |1\rangle_{\mathrm{BH}} |1\rangle_{\mathrm{GR}} \right) |0\rangle_{\mathrm{IN}} |0\rangle_{\mathrm{OUT}},
\end{align}
with normalization $|\alpha|^2 + |\beta|^2 = 1$.
The reduced states are :
\begin{align}
\rho_{\mathrm{BH,GR}}(\tau_1)&=\mathrm{Tr}_{\mathrm{IN,OUT}}(|\Psi\rangle\langle\Psi|_{\mathrm{BH,GR,IN,OUT}}(\tau_1))\cr
&=|\alpha|^2|00\rangle\langle00|+\alpha\beta^{*}|00\rangle\langle11|+\beta\alpha^{*}|11\rangle\langle00|+|\beta|^2|11\rangle\langle 11|,\\
\rho_{\mathrm{IN,OUT}}(\tau_1)&=|00\rangle\langle00|,
\end{align}
Tracing single-qubit subsystems gives
\begin{align}
    \rho_{\mathrm{BH}}(\tau_1) &= \mathrm{Tr}_{\mathrm{GR}}(\rho_{\mathrm{BH,GR}}(\tau_1)) = |\alpha|^2 |0\rangle\langle0| + |\beta|^2 |1\rangle\langle1|,\cr
     \rho_{\mathrm{GR}}(\tau_1)&= \mathrm{Tr}_{\mathrm{BH}}(\rho_{\mathrm{BH,GR}}(\tau_1)) = |\alpha|^2 |0\rangle\langle0| + |\beta|^2 |1\rangle\langle1|,
\end{align}
\begin{align}
    \rho_{\mathrm{IN}}(\tau_1) &= |0\rangle\langle0| ,\cr
     \rho_{\mathrm{OUT}}(\tau_1)&= |0\rangle\langle0|,
\end{align}
here we have suppressed the indices for the states for brevity.
The von Neumann entropy of BH and OUT subsystems are then:
\begin{align}
S_{\mathrm{BH}}(\tau_1) &= -|\alpha|^2 \log |\alpha|^2 - |\beta|^2 \log |\beta|^2,\\
S_{\mathrm{OUT}}(\tau_1) &= 0.
\end{align}
Mutual information is obtained by calculating the eigenvalues of the reduced density matrices : 
\begin{align}
    \mathcal{I}_{\mathrm{BH,GR}}(\tau_1) &= -2(|\alpha|^2 \log |\alpha|^2 + |\beta|^2 \log |\beta|^2),\cr
    \mathcal{I}_{\mathrm{IN:OUT}}(\tau_1) &= 0.
\end{align}
The partially transposed density matrix $\rho^{T_{\mathrm{GR}}}$ has eigenvalues \(\{|\alpha|^2,|\beta|^2,\pm|\alpha||\beta|\}\). The entanglement negativity of the subsystems are:
\begin{align}
    \mathcal{N}_{\mathrm{BH,GR}}(\tau_1) &= |\alpha||\beta|\cr
    \mathcal{N}_{\mathrm{IN,OUT}}(\tau_1) &= 0.
\end{align}

\subsection*{ IN-OUT entanglement at \boldmath$\tau_2$}

After $U_H$ on IN followed by CNOT$_{\mathrm{IN,OUT}}$:
\begin{align}
|\Psi(\tau_2)\rangle_{\mathrm{BH,GR,IN,OUT}} 
&= \left( \alpha |0\rangle_{\mathrm{BH}} |0\rangle_{\mathrm{GR}} + \beta |1\rangle_{\mathrm{BH}} |1\rangle_{\mathrm{GR}} \right) \otimes \left( \theta |0\rangle_{\mathrm{IN}} |0\rangle_{\mathrm{OUT}} + \eta |1\rangle_{\mathrm{IN}} |1\rangle_{\mathrm{OUT}} \right)\cr
\end{align}
Thus,
\begin{align}
\rho_{\mathrm{BH,GR}}(\tau_2)&=\rho_{\mathrm{BH,GR}}(\tau_1),\cr
\rho_{\mathrm{IN,OUT}}(\tau_2)&=\mathrm{Tr}_{\mathrm{BH,GR}}(|\Psi\rangle\langle\Psi|_{\mathrm{BH,GR,IN,OUT}}(\tau_2))\cr
&=|\theta|^2|00\rangle\langle00|+\theta\eta^{*}|00\rangle\langle11|+\eta\theta^{*}|11\rangle\langle00|+|\eta|^2|11\rangle\langle 11|,
\end{align}
The single-qubit reductions are
\begin{align}
    &\rho_{\mathrm{IN}}(\tau_2)  = |\theta|^2 |0\rangle\langle0| + |\eta|^2 |1\rangle\langle1|,\cr
    & \rho_{\mathrm{OUT}}(\tau_2)= |\theta|^2 |0\rangle\langle0| + |\eta|^2 |1\rangle\langle1|.
\end{align}

Entropies:
\begin{align}
&S_{\mathrm{BH}}(\tau_2) = S_{\mathrm{BH}}(\tau_1) \cr
 & S_{\mathrm{OUT}}(\tau_2) = -|\theta|^2 \log |\theta|^2 - |\eta|^2 \log |\eta|^2.
\end{align}

Mutual information:
\begin{align}
&\mathcal{I}_{\mathrm{BH,GR}}(\tau_2) =\mathcal{I}_{\mathrm{BH,GR}}(\tau_1) ,\cr
  &  \mathcal{I}_{\mathrm{IN,OUT}}(\tau_2) = -2(|\theta|^2 \log |\theta|^2 + |\eta|^2 \log |\eta|^2).
\end{align}
Eigenvalues of the partial transpose $\rho^{T_{\mathrm{OUT}}}$: \(\{|\theta|^2,|\eta|^2,\pm|\theta||\eta|\}\), so:
\begin{align}
&\mathcal{N}_{\mathrm{BH,GR}}(\tau_2) =\mathcal{N}_{\mathrm{BH,GR}}(\tau_1) ,\cr
&\mathcal{N}_{\mathrm{IN,OUT}}(\tau_2) = |\theta||\eta|.
\end{align}

\subsection*{Semi-causality violation \boldmath$\tau_3$}
The total state at $\tau_3$ is:
\begin{align}
|\Psi(\tau_3)\rangle_{\mathrm{BH,GR,IN,OUT}} 
&=\left(\theta |0\rangle_{\mathrm{BH}} |0\rangle_{\mathrm{GR}} 
- \eta \sin \sigma |1\rangle_{\mathrm{BH}} |0\rangle_{\mathrm{GR}} 
+ \eta \cos \sigma |0\rangle_{\mathrm{BH}} |1\rangle_{\mathrm{GR}}\right)\otimes \cr
&\,\,\,\,\left( \alpha |0\rangle_{\mathrm{IN}} |0\rangle_{\mathrm{OUT}} + \beta |1\rangle_{\mathrm{IN}} |1\rangle_{\mathrm{OUT}} \right).
\end{align}

The corresponding reduced density matrices are constructed as:
\begin{align}
\rho_{\mathrm{BH,GR}}(\tau_3)&=\mathrm{Tr}_{\mathrm{IN,OUT}}(|\Psi\rangle\langle\Psi|_{\mathrm{BH,GR,IN,OUT}}(\tau_3))\cr
&=|\theta|^2|00\rangle\langle00|+\theta\eta^{*}\cos \sigma|00\rangle\langle01|-\theta\eta^{*}\sin \sigma|00\rangle\langle10|+\eta\theta^{*}\cos \sigma|01\rangle\langle00|\cr
&+|\eta|^{2}\cos^2 \sigma|01\rangle\langle01|-|\eta|^{2}\cos \sigma\sin\sigma|01\rangle\langle10|-\eta\theta^{*}\sin \sigma|10\rangle\langle00|\cr
&-|\eta|^2\sin\sigma\cos\sigma|10\rangle\langle01|+|\eta|^2\sin^2 \sigma|10\rangle\langle10|,
\end{align}
\begin{align}
\rho_{\mathrm{IN,OUT}}(\tau_3)&=\mathrm{Tr}_{\mathrm{BH,GR}}(|\Psi\rangle\langle\Psi|_{\mathrm{BH,GR,IN,OUT}}(\tau_3))\cr
&=|\alpha|^2|00\rangle\langle00|+\alpha \beta^{*}|00\rangle\langle11|+ \beta\alpha^{*}|11\rangle\langle00|+| \beta|^2|11\rangle\langle 11|.
\end{align}
Similarly,
\begin{align}
\rho_{\mathrm{BH}}(\tau_3)&=\mathrm{Tr}_{\mathrm{GR}}[\rho_{\mathrm{BH,GR}}(\tau_3)]\cr
&=(|\theta|^2+|\eta|^{2}\cos^2\sigma)|0\rangle\langle0|-\theta\eta^{*}\sin\sigma|0\rangle\langle1|-\eta\theta^{*}\sin\sigma|1\rangle\langle0|+|\eta|^2\sin^2\sigma|1\rangle\langle1|,\cr
\end{align}
\begin{align}
\rho_{\mathrm{GR}}(\tau_3)&=\mathrm{Tr}_{\mathrm{BH}}[\rho_{\mathrm{BH,GR}}(\tau_3)]\cr
&=(|\theta|^2+|\eta|^{2}\sin^2\sigma)|0\rangle\langle0|+\theta\eta^{*}\cos\sigma|0\rangle\langle1|+\eta\theta^{*}\cos\sigma|1\rangle\langle0|+|\eta|^2\cos^2\sigma|1\rangle\langle1|,\cr
\end{align}
\begin{align}
    &\rho_{\mathrm{IN}}(\tau_3) = \mathrm{Tr}_{\mathrm{OUT}}(\rho_{\mathrm{IN,OUT}}(\tau_3))  = |\alpha|^2 |0\rangle\langle0| + |\beta|^2 |1\rangle\langle1|,\cr
    & \rho_{\mathrm{OUT}}(\tau_3)= \mathrm{Tr}_{\mathrm{IN}}(\rho_{\mathrm{IN,OUT}}(\tau_3)) = |\alpha|^2 |0\rangle\langle0| + |\beta|^2 |1\rangle\langle1|.
\end{align}
The reduced BH-GR state has eigenvalues:
\begin{align*}
\lambda_{\mathrm{BH},\mathrm{GR}}(\tau_3)&=1,\hspace{0.6cm} \lambda_{\mathrm{BH}}(\tau_3)=\lambda_{\mathrm{GR}}(\tau_3)=\frac{1}{2}(1\pm\sqrt{1-4|\eta|^4\cos^2\sigma\sin^2\sigma}),
\end{align*}
and  \(\rho_{\mathrm{IN,OUT}}(\tau_3)\) gives:
\begin{align*}
\lambda_{\mathrm{IN},\mathrm{OUT}}(\tau_3)&=1,\hspace{0.6cm} \lambda_{\mathrm{IN}}(\tau_3)=\lambda_{\mathrm{OUT}}(\tau_3)=|\alpha|^2 ,  |\beta|^2.
\end{align*}
Thus,
\begin{align}
&S_{\mathrm{BH}}(\tau_3) = -\log(|\eta|^2 \cos\sigma \sin\sigma) \cr
&- \frac{1}{2} \sqrt{1 - 4 |\eta|^4 \cos^2\sigma \sin^2\sigma} \log \left( \frac{1 + \sqrt{1 - 4 |\eta|^4 \cos^2\sigma \sin^2\sigma}}{1 - \sqrt{1 - 4 |\eta|^4\cos^2\sigma \sin^2\sigma}} \right),\\
& S_{\mathrm{OUT}}(\tau_3) = -|\alpha|^2 \log |\alpha|^2 -|\beta|^2\log |\beta|^2.
\end{align}
The mutual information are
\begin{align}
 &\mathcal{I}_{\mathrm{BH,GR}}(\tau_3) =-2\log(|\eta|^2 \cos\sigma \sin\sigma) \cr
&- \sqrt{1 - 4 |\eta|^4 \cos^2\sigma \sin^2\sigma} \log \left( \frac{1 + \sqrt{1 - 4 |\eta|^4 \cos^2\sigma \sin^2\sigma}}{1 - \sqrt{1 - 4 |\eta|^4\cos^2\sigma \sin^2\sigma}} \right),\\
& \mathcal{I}_{\mathrm{IN,OUT}}(\tau_3) = -2(|\alpha|^2 \log |\alpha|^2 +|\beta|^2\log |\beta|^2).
\end{align}

The partially transposed state is 
\begin{align*}
\rho_{\mathrm{BH,GR}}^{\mathrm{T_{GR}}}(\tau_3)
&=|\theta|^2|00\rangle\langle00|+\theta\eta^{*}\cos\sigma |01\rangle\langle00|-\theta\eta^{*}\sin\sigma|00\rangle\langle10|+\eta \theta^{*}\cos\sigma|00\rangle\langle01|\cr
&+|\eta|^{2}\cos^2\sigma |01\rangle\langle01|-|\eta|^{2}\cos\sigma \sin\sigma|00\rangle\langle11|-\eta\theta^{*}\sin\sigma|10\rangle\langle00|\cr
&-|\eta|^2\cos\sigma \sin\sigma|11\rangle\langle00|+|\eta|^2\sin^2\sigma|10\rangle\langle10|
\end{align*}
It has eigenvalues
 \begin{align*}
\lambda^{\mathrm{T_{GR}}}(\tau_3)&=\pm|\eta|^2\cos\sigma \sin\sigma, \frac{1}{2}[1
\pm\sqrt{1-4|\eta|^4\cos^2\sigma \sin^2\sigma}]
\end{align*}
giving
\begin{equation}
    \mathcal{N}_{\mathrm{BH,GR}}(\tau_3) =|\eta|^2\cos\sigma \sin\sigma.
\end{equation}
The partial transpose $\rho^{T_{\mathrm{OUT}}}$ has eigenvalues \(\{|\alpha|^2,|\beta|^2,\pm|\alpha||\beta|\}\), so:
\begin{align}
&\mathcal{N}_{\mathrm{IN,OUT}}(\tau_3) = |\alpha| |\beta|.
\end{align}
\subsection*{ Final state and entanglement at \boldmath$\tau_4$}

The final state at $\tau_4$ is:
\begin{align}
|\Psi(\tau_4)\rangle_{\mathrm{BH,GR,IN,OUT}} 
&= \big((|\theta|^2+|\eta|^2\cos\sigma) |0\rangle_{\mathrm{BH}} |0\rangle_{\mathrm{GR}} 
+(\theta^* \eta\cos\sigma-\theta \eta^*)|0\rangle_{\mathrm{BH}} |1\rangle_{\mathrm{GR}} \cr
 &- \theta^*\eta\sin\sigma |1\rangle_{\mathrm{BH}} |0\rangle_{\mathrm{GR}} 
 +|\eta|^2\sin\sigma |1\rangle_{\mathrm{BH}} |1\rangle_{\mathrm{GR}}\big)\otimes \cr
 &\big(\alpha|0\rangle_{\mathrm{IN}}|0\rangle_{\mathrm{OUT}}+\beta\cos\sigma|0\rangle_{\mathrm{IN}}|1\rangle_{\mathrm{OUT}}-\beta\sin\sigma|1\rangle_{\mathrm{IN}}|0\rangle_{\mathrm{OUT}}\big).
\end{align}

 Since the BH-GR sector undergoes a local unitary on GR,
 \begin{align}
\rho_{\mathrm{BH,GR}}(\tau_4)&=\mathrm{Tr}_{\mathrm{IN,OUT}}(|\Psi\rangle\langle\Psi|_{\mathrm{BH,GR,IN,OUT}}(\tau_4))\cr
&=\rho_{\mathrm{BH,GR}}(\tau_3)
\end{align}
The reduced  IN-OUT state is
\begin{align}
\rho_{\mathrm{IN,OUT}}(\tau_4)&=\mathrm{Tr}_{\mathrm{BH,GR}}(|\Psi\rangle\langle\Psi|_{\mathrm{BH,GR,IN,OUT}}(\tau_4))\cr
&=|\alpha|^2|00\rangle\langle00|+\alpha\beta^{*}\cos\sigma|00\rangle\langle01|-\alpha\beta^{*}\sin\sigma|00\rangle\langle10|+\beta\alpha^{*}\cos\sigma|01\rangle\langle00|\cr
&+|\beta|^{2}\cos^2\sigma|01\rangle\langle01|-|\beta|^{2}\cos\sigma\sin\sigma|01\rangle\langle10|-\beta\alpha^{*}\sin\sigma|10\rangle\langle00|\cr
&-|\beta|^2\cos\sigma\sin\sigma|10\rangle\langle01|+|\beta|^2\sin^2\sigma|10\rangle\langle10|.
\end{align}

Then we compute the single qubit states $\rho_{\mathrm{BH}}(\tau_4)$, $\rho_{\mathrm{OUT}}(\tau_4)$ and the partially transposed state and derive: 
\begin{align}
&S_{\mathrm{BH}}(\tau_4)=S_{\mathrm{BH}}(\tau_3),\cr
&\mathcal{I}_{\mathrm{BH}, \mathrm{GR}}(\tau_4) = \mathcal{I}_{\mathrm{BH}, \mathrm{GR}}(\tau_3),\cr
&\mathcal{N}_{\mathrm{BH}, \mathrm{GR}}(\tau_4) = \mathcal{N}_{\mathrm{BH}, \mathrm{GR}}(\tau_3) ,
\end{align}
\begin{align}
S_{\mathrm{OUT}}(\tau_4)
&=-\ln(|\beta|^2\cos\sigma\sin\sigma)\cr
&-\frac{1}{2}\sqrt{1-4|\beta|^4\cos^2\sigma\sin^2\sigma}\ln\frac{1+\sqrt{1-4|\beta|^4\cos^2\sigma\sin^2\sigma}}{1-\sqrt{1-4|\beta|^4\cos^2\sigma\sin^2\sigma}},
\end{align}
\begin{align}
\mathcal{I}_{\mathrm{IN,OUT}}(\tau_4)
&=-2\ln(|\beta|^2\cos\sigma\sin\sigma)\cr
&-\sqrt{1-4|\beta|^4\cos^2\sigma\sin^2\sigma}\ln\frac{1+\sqrt{1-4|\beta|^4\cos^2\sigma\sin^2\sigma}}{1-\sqrt{1-4|\beta|^4\cos^2\sigma\sin^2\sigma}},
\end{align}
\begin{equation}
    \mathcal{N}_{\mathrm{IN,OUT}}(\tau_4) = |\beta|^2 \cos\sigma \sin\sigma.
\end{equation}

\begingroup
\setlength{\emergencystretch}{2em}
\sloppy
\bibliographystyle{apsrev4-2}
\bibliography{self}
\endgroup

\end{document}